\pdfoutput=1
\documentclass[10pt]{iopart}

\usepackage{iopams}
\usepackage{setstack}
\usepackage{graphicx}

\begin{document}

\title[Finite-$\beta$ hybrid configurations by planar dipole-field coils]{Optimized finite-$\beta$ tokamak-stellarator hybrid configurations achieved by planar dipole-field coils}

\author{Yihui Liang$^1$, Hengqian Liu$^2$, Guodong Yu$^{2,*}$, Zhenyu Zhou$^1$, Caoxiang Zhu$^{2}$, Yao Zhou$^{1,*}$}

\address{$^1$ School of Physics and Astronomy, Institute of Natural Sciences, and MOE-LSC, Shanghai Jiao Tong University, Shanghai 200240, China}
\address{$^2$ School of Nuclear Science and Technology, University of Science and Technology of China, Hefei 230027, China}

\eads{\mailto{gdyu@ustc.edu.cn,yao.zhou@sjtu.edu.cn}}

\begin{abstract}
Tokamak--stellarator hybrids seek to combine tokamak-like compactness and confinement with stellarator-like externally generated rotational transform and steady-state operation.
In this work, we build on the recent tokamak--stellarator hybrid study using planar dipole-field coils (PDCs) [Yu et al., arXiv:2605.03599], in which the fixed-position, programmable coils on an axisymmetric winding surface generate flexible three-dimensional shaping fields.
Using single-stage free-boundary optimization of coil currents and plasma-equilibrium parameters, we construct vacuum and finite-$\beta$ configurations.
The vacuum cases show controllable external transform and magnetic well.
The finite-$\beta$ cases accommodate various density, temperature, and pressure profiles, producing quasi-axisymmetric (QA) equilibria with self-consistent bootstrap current, favorable Mercier stability, and reduced demand for external current drive.
Re-optimization enables $\beta$ ramp-up and access to different field-period QA branches with moderate coil-current changes.
At large rotational transform, a toroidally omnigenous (TO)-like configuration exhibits more favorable infinite-$n$ ideal-ballooning behavior than a QA reference with matched profiles, even though ballooning stability is not directly optimized for.
These results demonstrate that PDCs provide a flexible platform for achieving optimized finite-$\beta$ hybrid configurations.
\end{abstract}

\vspace{2pc}

\maketitle

\ioptwocol

\section{Introduction}
\label{sec:introduction}

Tokamaks and stellarators are two leading concepts for toroidal magnetic confinement fusion \cite{Freidberg2007,Boozer2005,ImbertGerard2020}. 
Tokamaks provide compact geometry, large volumes of nested magnetic surfaces, and favorable confinement properties, but require a large toroidal plasma current, which introduces current-driven instabilities and complicates steady-state operation \cite{Boozer2005,Buttery2021}. 
Stellarators generate rotational transform mainly with external three-dimensional coils and are intrinsically compatible with steady-state operation, but optimized stellarator coils are usually geometrically complex \cite{ImbertGerard2020,Wechsung2022}. 
Tokamak--stellarator hybrids seek to combine these complementary advantages by sharing the total rotational transform between external coils and plasma current.

The degree of hybridization may be described by the external rotational-transform fraction $\eta_{\rm ERT}$, with $\eta_{\rm ERT}=0$ corresponding to the tokamak limit and $\eta_{\rm ERT}=1$ to the current-free stellarator limit. 
Intermediate values have been explored in several hybrid concepts, including TOKASTAR, low-aspect-ratio stellarator/tokamak hybrids, the Compact Toroidal Hybrid (CTH), compact quasi-axisymmetric (QA) hybrids, and QA tokamak concepts for steady-state operation \cite{Yamazaki1985,Moroz1996,Spong1998,Okano2010,Hartwell2017,Ku2009,Henneberg2024,Schuett2024,Liang2025NF}. 
Experiments on CTH have shown that externally generated rotational transform can mitigate current-driven MHD instabilities and help avoid disruptions \cite{Pandya2015}. 
In advanced tokamak scenarios, externally generated rotational transform may also reduce the external current-drive requirement by replacing part of the plasma-current contribution to rotational transform, with the remaining part supplied by the bootstrap current \cite{Buttery2021,Polevoi2020,Liang2025NF}. 
More generally, hybrid configurations offer a practical route to access multiple magnetic configurations in one device, provided that the required three-dimensional fields can be varied without hardware redesign.

This requirement motivates coil systems that are simple, modular, and current-programmable. 
Conventional optimized stellarators rely on complex modular coils, whereas alternative concepts based on saddle coils, permanent magnets, and planar coils aim to reduce engineering complexity while retaining three-dimensional shaping capability \cite{Liang2025NF,Henneberg2024,Helander2020,Gates2025PlanarCoils}. 
Recent planar-coil stellarator studies are especially relevant: large plasma-encircling coils can provide the main field, while smaller dipole-field coils, arranged on a non-axisymmetric winding surface, control the remaining non-axisymmetric field components \cite{Gates2025PlanarCoils,Wu2025Eos}. 
This architecture has been further developed toward the reactor-scale Helios design \cite{Swanson2025Helios}, and the Canis experiment has demonstrated high-temperature-superconducting planar shaping coils with closed-loop control of stellarator-relevant fields \cite{Nash2025Canis}. 

This approach has recently been generalized to a programmable stellarator--tokamak hybrid, which shows that a fixed set of simple planar dipole-field coils (PDCs) placed on an axisymmetric winding surface can generate a large magnetic-configuration space by varying coil currents, including optimized QA, quasi-helical-symmetric (QH), and quasi-isodynamic (QI) stellarator configurations, as well as tokamak-relevant three-dimensional perturbations \cite{Yu2026PDC}. 
A more recent study further examined the feasibility of this concept on the HBT-EP experiment \cite{Halpern2026}.
These result demonstrate that magnetic-configuration exploration can be partially decoupled from hardware redesign. 
The present work is complementary:
we extend the PDC framework to construct finite-$\beta$ equilibria with self-consistent current profiles and further explore variations in profiles, $\beta$, field period, and QA/TO character.

We adopt a single-stage free-boundary optimization framework for PDC-supported hybrid configurations. 
The PDCs are arranged on an axisymmetric winding surface with fixed positions and programmable currents. 
The PDC currents, poloidal-field (PF) currents, and selected plasma-equilibrium parameters are optimized simultaneously, allowing different three-dimensional shaping equilibria to be generated within the same device concept.
We first obtain vacuum QA configurations with controllable external rotational transform and magnetic well. 
We then construct finite-$\beta$ QA hybrid equilibria with prescribed density, temperature, and pressure profiles, including representative cases with self-consistent bootstrap current and favorable Mercier stability over most of the plasma volume. 
These cases are relevant to steady-state operation because the external coils supply part of the rotational transform that would otherwise need to be produced by external current drive. 
By re-optimizing the current distributions and equilibrium parameters, we further demonstrate $\beta$ ramp-up and field-period variation.
We also explore a toroidal omnigenity (TO)-like extension at large rotational transform (large-$\iota$) and compare QA and TO-like configurations with matched prescribed pressure and current-density profiles after Mercier-stability optimization. 
We find that the TO-like configuration has better ideal-ballooning stability than the QA reference, even though ballooning stability is not included in the optimization objective.
These results show that the PDC framework can support a family of physically distinct finite-$\beta$ hybrid configurations within one free-boundary optimization platform, which is useful for experimentally relevant configuration scans and for developing hybrid devices that can test multiple magnetic-confinement concepts without hardware redesign.

The rest of this paper is organized as follows. 
Section~\ref{sec:method} introduces the single-stage optimization method, the objective function, and the configuration labels used throughout the paper. 
Section~\ref{sec:vacuum} presents the vacuum QA configurations and discusses the control of external rotational transform and magnetic well. 
Section~\ref{sec:finite_beta} focuses on finite-$\beta$ hybrid configurations, including profile variation, self-consistent bootstrap current, $\beta$ ramp-up, and field-period variation. 
Section~\ref{sec:TO} compares large-$\iota$ QA and TO-like configurations and analyzes their ideal-ballooning stability. 
Section~\ref{sec:summary} summarizes the main conclusions and discusses future directions.

\section{Optimization framework}
\label{sec:method}

The optimized configurations are obtained using a single-stage free-boundary optimization method. 
In a conventional two-stage procedure, a fixed-boundary equilibrium is first optimized and coils are then designed to reproduce it. 
This procedure provides useful initial guesses, but the resulting free-boundary equilibrium may deviate from the target once coil constraints are imposed. 
In this work, the two-stage results are therefore used only to initialize the equilibrium and coil currents, and the final configurations are obtained by subsequent single-stage free-boundary optimization.

The coil set includes toroidal-field (TF) coils, poloidal-field (PF) coils, and PDCs. 
The TF coil currents and all coil positions are fixed. 
The optimized variables include the PF coil currents, the PDC currents, and selected plasma-equilibrium parameters, including current-profile coefficients. 
The PDCs are distributed on an axisymmetric winding surface with a resolution of $24\times 16$ in the toroidal and poloidal directions, respectively. 
With this fixed layout, three-dimensional shaping is controlled through the programmed current distribution of the planar coils, providing an effective representation of saddle-coil-like shaping fields without changing the coil geometry.

The initial fixed-boundary equilibria and coil-current distributions are generated with SIMSOPT and FOCUS \cite{Landreman2021SIMSOPT,Zhu2018FOCUS}. 
The single-stage optimization is then carried out using a modified version of SIMSOPT that couples the optimization variables directly to free-boundary VMEC calculations \cite{Landreman2021SIMSOPT,Hirshman1983VMEC}.
DESC is used for further equilibrium optimization and post-processing \cite{Dudt2020DESC}, and the TO-like case is constructed using the OOPS method \cite{Liu2025OOPS,Dudt2024Omnigenity}. 
After optimization, STELLOPT-based tools including XBOOZ\_XFORM and NEO are used to evaluate Boozer spectra, neoclassical transport, and bootstrap-current consistency \cite{Boozer1981Coordinates,Sanchez2000BOOZ,Nemov1999NEO}. 
DESC-based diagnostics are further used to evaluate Mercier stability and ideal-ballooning stability, namely the ideal-ballooning metrics shown in Section \ref{sec:TO}.

The total objective function is written as
\begin{eqnarray}
f_{\rm total}=\omega_{\bar{A}} f_{\bar{A}}+\omega_{\rm ERT} f_{\rm ERT}+\omega_{\rm Well/Merc} f_{\rm Well/Merc} \nonumber \\
\qquad +\omega_{\rm BS} f_{\rm BS}+\omega_{\rm QA/TO} f_{\rm QA/TO} +\sum_{i}\omega^{\rm coil}_{i} f^{\rm coil}_{i},
\label{eq:objective_total}
\end{eqnarray}
where the $\omega$'s are prescribed weights. 
The weights also absorb the relative normalization and units of the different objective terms.

The objective terms are defined as follows.

\begin{itemize}

\item \textit{Aspect ratio.}
The compactness of the plasma is controlled using the averaged aspect ratio defined in the same way as in \cite{Liang2025NF},
\begin{eqnarray}
\bar{A}=\frac{\bar{R}}{\bar{a}}, \quad
\bar{R}=\frac{V}{2\pi\bar{S}}, \quad
\bar{a}=\left(\frac{\bar{S}}{\pi}\right)^{1/2}.
\label{eq:averaged_aspect_ratio}
\end{eqnarray}
Here $V$ is the plasma volume and $\bar{S}$ is the average area of the toroidal cross section. 
The corresponding objective is
\begin{eqnarray}
f_{\bar{A}}=(\bar{A}-\bar{A}^{\ast})^2,
\label{eq:f_aspect}
\end{eqnarray}
where $\bar{A}^{\ast}$ is the target value.

\item \textit{External rotational transform.}
The external rotational transform (ERT) is defined as the part of the rotational transform supplied by the external coils. 
It is evaluated from the final equilibrium by subtracting the plasma-current contribution,
\begin{equation}
\iota_{\rm ERT}(s)=\iota(s)-\iota_{\rm cur}(s),
\label{eq:iota_ert}
\end{equation}
where $\iota_{\rm cur}(s)$ is the rotational transform generated by the plasma current. 
In contrast to the diagnostic used in Ref.~\cite{Liang2025NF}, the ERT here is not obtained by recomputing a vacuum field with the same boundary. 
For vacuum configurations, $\iota_{\rm cur}(s)=0$ and therefore $\iota_{\rm ERT}(s)=\iota(s)$. 
The averaged ERT is defined as
\begin{equation}
\bar{\iota}_{\rm ERT}=\int_0^1 \iota_{\rm ERT}(s)\,ds .
\label{eq:iota_ert_average}
\end{equation} 
When an ERT target is imposed, the objective is
\begin{equation}
f_{\rm ERT}=(\bar{\iota}_{\rm ERT}-\bar{\iota}_{\rm ERT}^{\ast})^2 .
\label{eq:f_ert}
\end{equation}

\item \textit{Magnetic well.}
The magnetic-well parameter used in the DESC diagnostics is defined as \cite{Dudt2020DESC,LandremanJorge2020}
\begin{equation}
W(s)=\frac{V}{\left\langle B^2\right\rangle}\frac{d}{dV}\left\langle2\mu_0 p+B^2\right\rangle .
\label{eq:magnetic_well}
\end{equation}
Here $V$ is the plasma volume enclosed by a flux surface and $\langle \cdots \rangle$ denotes a flux-surface average. 
A positive value of $W$ is favorable for interchange stability and used as a stability proxy mainly for vacuum field optimization.
In the optimization, the magnetic-well objective is imposed on a selected flux surface $s=s_W$, not as a volume average.
When it is used as a lower-bound target, the penalty is
\begin{equation}
f_{\rm Well}=\left[\min\left(0,W(s_W)-W^{\ast}\right)\right]^2.
\label{eq:f_well}
\end{equation}
Here $W^{\ast}$ is the prescribed lower bound.

\item \textit{Mercier stability.}
The Mercier criterion is used as a metric for interchange stability of finite-$\beta$ equilibria \cite{Mercier1964,LandremanJorge2020}.
We denote the DESC/VMEC Mercier value by $D_{\rm Merc}(s)$.
Mercier stability corresponds to
\begin{equation}
D_{\rm Merc}(s)>0 .
\label{eq:mercier_stable}
\end{equation}
As with the magnetic-well term, the optimization is applied to a selected flux surface $s=s_M$.
The lower-bound penalty is
\begin{equation}
f_{\rm Merc}=\left[\min\left(0,D_{\rm Merc}(s_M)-D_{\rm Merc}^{\ast}\right)\right]^2.
\label{eq:f_merc}
\end{equation}
Here $D_{\rm Merc}^{\ast}$ is the target lower bound.

\item \textit{Bootstrap-current consistency.}
For the finite-$\beta$ QA hybrid cases, the toroidal-current profile is optimized so that the equilibrium parallel current is consistent with the neoclassical bootstrap-current prediction.
The bootstrap current is calculated using the Redl formulae \cite{Redl2021Bootstrap}.
The mismatch objective is
\begin{equation}
f_{\rm BS}=\frac
{\int ds[\langle\bf{J}\cdot \bf{B}\rangle_{\mathrm{eq}}-\langle\bf{J}\cdot \bf{B}\rangle_{\mathrm{Redl}}]^2}
{\int ds[\langle\bf{J}\cdot \bf{B}\rangle_{\mathrm{eq}}+\langle\bf{J}\cdot \bf{B}\rangle_{\mathrm{Redl}}]^2}.
\label{eq:f_bs}
\end{equation}
Here $\langle{\bf J}\cdot{\bf B}\rangle_{\rm eq}$ is obtained from the optimized equilibrium, while $\langle{\bf J}\cdot{\bf B}\rangle_{\rm Redl}$ is calculated from the prescribed density and temperature profiles.
The equilibrium calculation uses the toroidal current profile $dI(s)/ds$ as an input, and the parallel-current profile is evaluated in post-processing.

\item \textit{Quasi-axisymmetry.}
The magnetic-field strength in Boozer coordinates is expanded as
\begin{equation}
B(s,\theta_B,\zeta_B)=\sum_{m,n}B_{mn}(s)\cos\left(m\theta_B-n_{\rm fp}n\zeta_B\right).
\label{eq:boozer_expansion}
\end{equation}
For QA configurations, $B$ should be independent of the Boozer toroidal angle.
The QA-breaking objective is evaluated on a selected flux surface $s=s_{\rm QA}$ and is written as
\begin{equation}
f_{\rm QA}=\sum_{m}||B_{m,n\neq0}(s_{\rm QA})||/B_{\rm 0,0},
\label{eq:f_qa}
\end{equation}
where $||\cdot||$ denotes $L^2$-norm.

\item \textit{Toroidal omnigenity.}
The TO-like case is optimized using the OOPS method.
OOPS introduces a coordinate mapping from a generalized Cary-Shasharina coordinate system to Boozer coordinates, so that an omnigenous field can be represented by straight $B$-contours in the mapped coordinates \cite{Liu2025OOPS,Dudt2024Omnigenity}.
Let $(\vartheta,\varphi)$ denote the mapped coordinates used for the TO target.
The field strength is expanded as
\begin{equation}
B(s,\vartheta,\varphi)=\sum_{\mu,\nu}\widetilde{B}_{\mu\nu}(s)\cos\left(\mu\vartheta-n_{\rm fp}\nu\varphi\right) .
\label{eq:oops_expansion}
\end{equation}
For a TO field, the field-strength contours close toroidally in the mapped coordinate system.
The TO-breaking level is reduced by suppressing the corresponding asymmetric modes.
In the present optimization, the TO objective is imposed on a selected flux surface $s=s_{\rm TO}$,
\begin{equation}
f_{\rm TO}=\sum_{\mu}||\widetilde{B}_{\mu,\nu\neq0}(s_{\rm TO}) ||/B_{\rm 0,0}.
\label{eq:f_to}
\end{equation}
This term is used only to construct the TO-like large-rotational-transform comparison case.

\item \textit{Coil-related constraints.}
The coil-related term represents engineering and regularization objectives for the coil system. 
In general, $f_i^{\rm coil}$ may include constraints on coil currents, electromagnetic forces, mutual inductance, self-inductance, or other engineering quantities. 
In this work, we mainly use current-related penalties to regularize the PDC currents. 
For example, the maximum-current penalty is written as
\begin{equation}
f^{\rm coil}_{\max}=\sum_k\left[\max\left(0,\,|I_k|-I_{\max}^{\ast}\right)\right]^2 .
\label{eq:f_coil_max}
\end{equation}
Here $I_k$ is the current in the $k$-th PDC and $I_{\max}^{\ast}$ is the prescribed upper bound. 
The average-current penalty is
\begin{equation}
f^{\rm coil}_{\rm ave}=\left(\left\langle|I_{\rm PDC}|\right\rangle-I_{\rm ave}^{\ast}\right)^2 .
\label{eq:f_coil_ave}
\end{equation}
Here $\langle |I_{\rm PDC}|\rangle$ is the average absolute current of the PDCs and $I_{\rm ave}^{\ast}$ is the target value. 
These examples illustrate the current regularization used in the present optimizations, while the same framework can accommodate additional coil-related metrics when more detailed engineering constraints are included.

\end{itemize}

In this work we present representative examples of the hybrid configurations obtained, which are summarized in Table~\ref{tab:case_summary}. 
The table provides a compact reference for the case labels used throughout the paper. 
More detailed construction, diagnostics, and physical interpretation are given in the corresponding result sections and in Table~\ref{tab:diagnostic_summary}.

\begin{table*}[t]
\centering
\caption{
Summary of the configurations considered in this work.
The table gives the case labels, configuration type, field period, averaged aspect ratio and major radius, volume-averaged magnetic field strength and $\beta$, prescribed profile family, and the section in which each case is discussed.
For the vacuum cases, the prefix V denotes vacuum and the final number in the label gives the approximate volume-averaged external rotational transform.
For the finite-$\beta$ cases, C, I, and X denote cubic, ITER-like, and NCSX-like profile families, and the final number gives the approximate percentage value of $\langle\beta\rangle$.
QA denotes quasi-axisymmetry, while TO denotes the toroidally omnigenous-oriented case.
The postfix Li means large rotational transform.
}

\label{tab:case_summary}
\begin{tabular}{ccccccccc}
\br
Case 
& Description 
& $n_{\rm fp}$ 
& $\bar{A}$
& $\bar{R}$ [m]
& $\langle B \rangle$ [T]
& $\langle\beta\rangle$ [\%]
& Profiles 
& Section \\
\mr
VQA2-i12 
& vacuum QA 
& $2$ 
& $3.97$ 
& $1.08$
& $0.94$
& $0$ 
& -- 
& Sec.~\ref{sec:vacuum} \\

VQA2-i20 
& vacuum QA
& $2$ 
& $3.99$ 
& $1.09$
& $0.93$
& $0$ 
& -- 
& Sec.~\ref{sec:vacuum} \\

\mr
QA2-C5 
& finite-$\beta$ QA 
& $2$ 
& $3.98$ 
& $1.08$
& $0.97$
& $4.71$ 
& cubic 
& Sec.~\ref{sec:high_beta} \\

QA2-I4 
& finite-$\beta$ QA 
& $2$ 
& $3.98$ 
& $1.08$
& $0.97$
& $3.83$ 
& ITER-like 
& Sec.~\ref{sec:high_beta} \\

QA2-X1 
& finite-$\beta$ QA  
& $2$ 
& $3.99$ 
& $1.07$
& $0.98$
& $1.32$ 
& NCSX-like 
& Sec.~\ref{sec:beta_ramp} \\

QA2-X3 
& finite-$\beta$ QA
& $2$ 
& $3.99$ 
& $1.08$
& $0.98$
& $3.36$ 
& NCSX-like 
& Sec.~\ref{sec:beta_ramp} \\

QA2-X5 
& finite-$\beta$ QA
& $2$ 
& $3.98$
& $1.08$
& $0.97$
& $5.44$ 
& NCSX-like 
& Secs.~\ref{sec:high_beta}, \ref{sec:beta_ramp} \\

\mr
QA3-X3 
& finite-$\beta$ QA
& $3$ 
& $5.70$ 
& $1.03$
& $1.08$
& $2.73$ 
& NCSX-like 
& Sec.~\ref{sec:nfp_variation} \\

QA4-X3 
& finite-$\beta$ QA
& $4$ 
& $4.86$ 
& $1.02$
& $1.14$
& $2.46$ 
& NCSX-like 
& Sec.~\ref{sec:nfp_variation} \\

\mr
QA2-X3-Li
& large-$\iota$ QA
& $2$ 
& $4.00$ 
& $1.09$
& $0.94$
& $3.56$ 
& NCSX-like 
& Sec.~\ref{sec:TO} \\

TO2-X3-Li 
& large-$\iota$ TO
& $2$ 
& $4.00$ 
& $1.07$
& $0.94$
& $3.54$ 
& NCSX-like 
& Sec.~\ref{sec:TO} \\

\br
\end{tabular}
\end{table*}

\begin{table*}[t]
\centering
\caption{
Selected diagnostic quantities for the optimized configurations. 
Here $\iota^{\rm LCFS}$ is the rotational transform at the LCFS, and $\eta_{\rm ERT}$ is the external-rotational-transform fraction. 
The bootstrap-current fraction $\eta_{\rm BS}$ is evaluated after optimization. 
A dash indicates that the corresponding quantity is not applicable or not used as a diagnostic for that case.
The symmetry-breaking level $B^{\rm err-LCFS}_{\rm QA/TO}$ denotes the QA-breaking level for QA configurations and the TO-breaking level for the TO-like configuration on the LCFS [Eqs. (\ref{eq:f_qa}) and (\ref{eq:f_to})]. 
The quantity $\epsilon_{\rm eff}^{\rm 3/2-LCFS}$ is the effective-ripple metric evaluated at the LCFS.   
The last two columns give the maximum and averaged PDC currents normalized by the single TF-coil current.
}
\label{tab:diagnostic_summary}
\begin{tabular}{ccccccccc}
\br
Case 
& $\iota^{\rm LCFS}$
& $\eta_{\rm ERT}$ 
& $\eta_{\rm BS}$ 
& $B^{\rm err-LCFS}_{\rm QA/TO}$ 
& $\epsilon_{\rm eff}^{\rm 3/2-LCFS}$ 
& $I_{\rm PDC}^{\max}/I_{\rm TF}$ 
& $I_{\rm PDC}^{\rm ave}/I_{\rm TF}$ \\
\mr

VQA2-i12 
& $0.11$
& $1.00$ 
& -- 
& $2.30\times10^{-2}$
& $3.50\times10^{-3}$
& $4.21$ 
& $0.65$ \\

VQA2-i20 
& $0.19$
& $1.00$ 
& -- 
& $3.44\times10^{-2}$
& $8.43\times10^{-3}$
& $6.56$ 
& $0.93$ \\

\mr
QA2-C5 
& $0.57$
& $0.22$ 
& $0.96$ 
& $1.95\times10^{-2}$
& $1.13\times10^{-2}$
& $3.78$ 
& $0.58$ \\

QA2-I4 
& $0.43$
& $0.17$ 
& $0.96$ 
& $1.44\times10^{-2}$
& $1.44\times10^{-3}$
& $3.92$ 
& $0.59$ \\

QA2-X1 
& $0.36$
& $0.41$ 
& $1.00$ 
& $1.09\times10^{-2}$
& $2.40\times10^{-4}$
& $4.03$ 
& $0.60$ \\

QA2-X3 
& $0.51$
& $0.29$ 
& $1.00$ 
& $1.33\times10^{-2}$
& $6.14\times10^{-4}$
& $3.89$ 
& $0.58$ \\

QA2-X5 
& $0.61$
& $0.26$ 
& $1.00$ 
& $1.39\times10^{-2}$
& $1.48\times10^{-3}$
& $3.81$ 
& $0.58$ \\

\mr
QA3-X3 
& $0.63$
& $0.17$ 
& $1.00$ 
& $5.55\times10^{-3}$
& $2.16\times10^{-4}$
& $3.72$ 
& $0.64$ \\

QA4-X3 
& $0.56$
& $0.19$ 
& $1.00$ 
& $2.78\times10^{-2}$
& $4.20\times10^{-4}$
& $5.59$ 
& $0.67$ \\

\mr
QA2-X3-Li
& $0.86$
& $0.21$ 
& $0.37$
& $8.45\times10^{-3}$
& $5.00\times10^{-4}$
& $5.70$ 
& $0.84$ \\

TO2-X3-Li 
& $0.85$
& $0.22$ 
& $0.36$
& $4.24\times10^{-3}$ 
& $9.77\times10^{-5}$
& $4.46$ 
& $0.75$ \\

\br
\end{tabular}
\end{table*}

\section{Vacuum QA configurations}
\label{sec:vacuum}

We first optimize vacuum QA configurations using the fixed-position PDC array. 
These cases test the ability of the planar-coil current distribution to generate external rotational transform, QA shaping, and magnetic well before finite-$\beta$ and plasma-current effects are introduced. 
All vacuum configurations considered here have $n_{\rm fp}=2$ and stellarator symmetry. 
The active targets are the averaged aspect ratio, ERT, QA breaking, magnetic well, and coil-current constraints, corresponding to nonzero weights $\omega_{\bar{A}}$, $\omega_{\rm ERT}$, $\omega_{\rm QA}$, $\omega_{\rm Well}$, and the relevant $\omega_i^{\rm coil}$.

\begin{figure*}[htbp]
    \centering
    \includegraphics[width=1.0\linewidth]{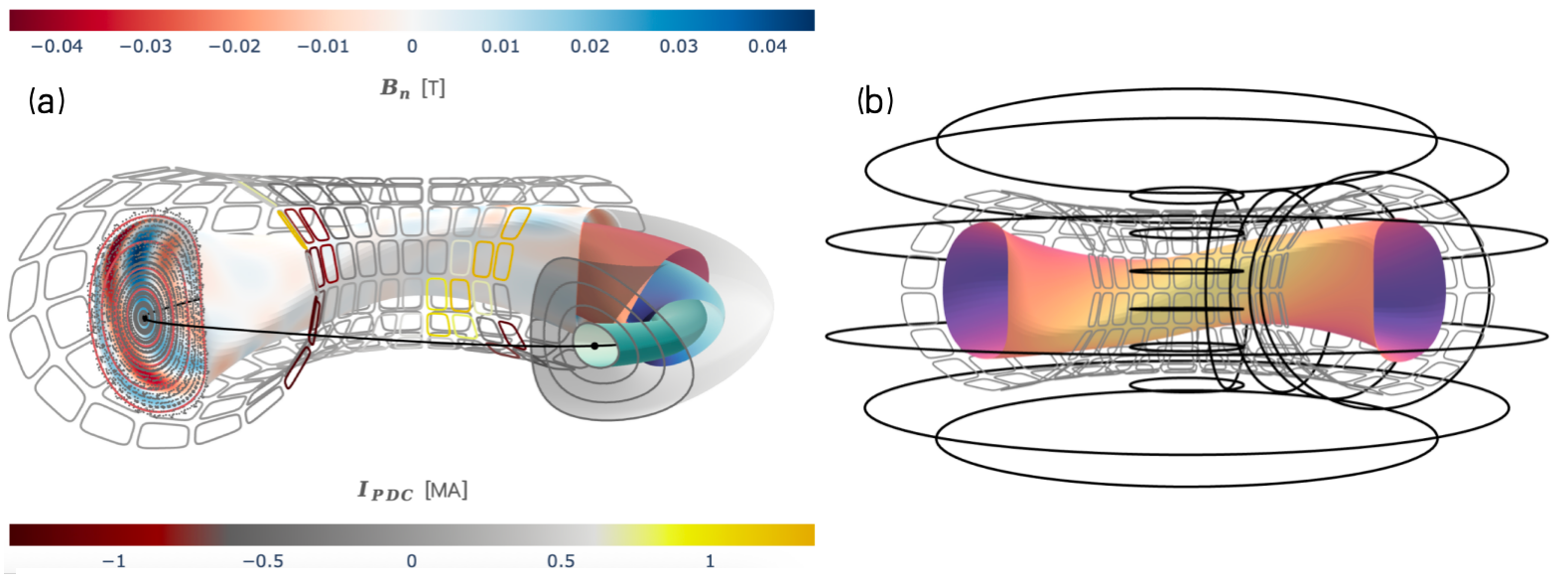}
    \caption{
    Representative PDC-supported vacuum configuration VQA2-i12 and coil layout. 
    (a) Distribution of the normal magnetic field $B_n$ on the LCFS and the corresponding PDC current distribution on the axisymmetric winding surface. 
    The upper colorbar denotes $B_n$, while the lower colorbar denotes the PDC current $I_{\rm PDC}$. 
    (b) Three-dimensional view of the plasma, PDC array, and external axisymmetric coil system. 
    The black solid curves denote the TF and PF coils; the three equal-size central pairs represent the effective PF coils associated with the central solenoid in the steady-state operating mode. 
    The grey line segments denote the PDCs.
    }
    \label{fig1}
\end{figure*}

Figure~\ref{fig1} shows the representative vacuum configuration VQA2-i12. 
The normal magnetic field $B_{\rm n}$ on the target boundary is shown together with the optimized PDC current distribution. 
In the vacuum calculation, $B_{\rm n}$ is generated entirely by the external coils; in the finite-$\beta$ equilibria discussed later, the corresponding normal field also contains contributions associated with plasma current. 
The upper colorbar in Fig.~\ref{fig1} corresponds to $B_{\rm n}$, while the lower colorbar gives the PDC currents. 
The optimized current pattern contains both positive and negative currents, indicating that the effective three-dimensional shaping is produced collectively by the planar-coil array. 
The Poincar\'e plot confirms that the resulting vacuum field has well-defined nested flux surfaces.

\begin{figure*}[htbp]
    \centering
    \includegraphics[width=1.0\linewidth]{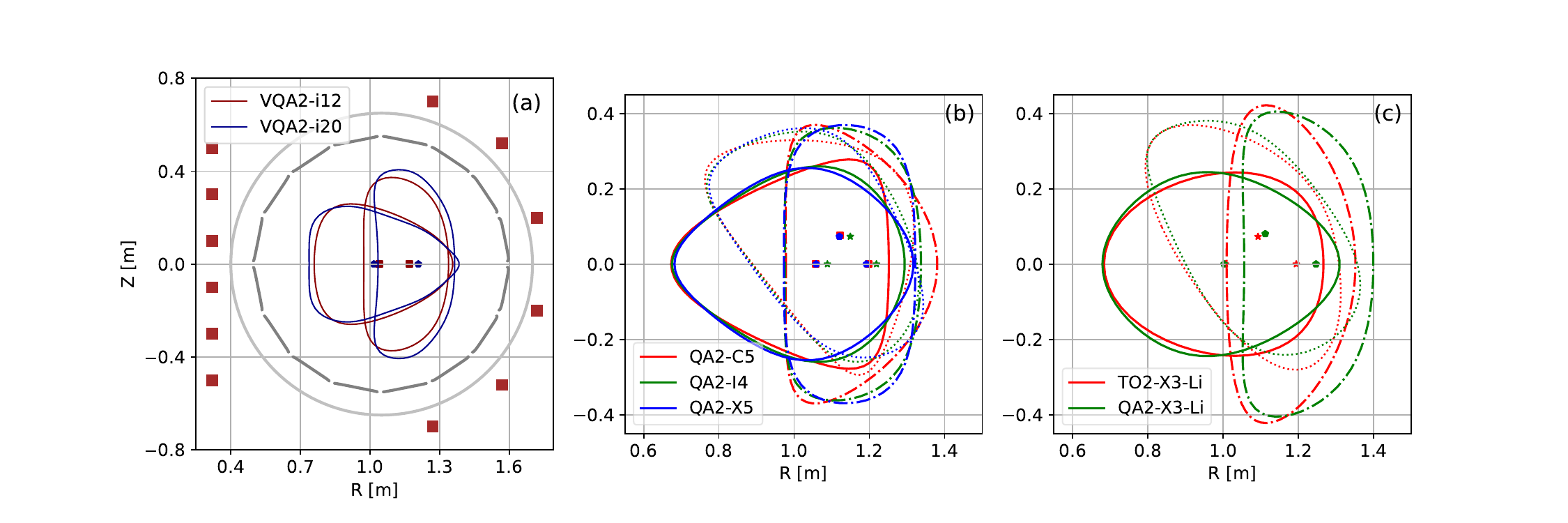}
    \caption{
    Overview of representative optimized configurations. 
    (a) Vacuum QA configurations with different external rotational transform and positive magnetic well. 
    The projections of the coils are also shown: PDCs are plotted as grey line segments, TF coils as light grey solid curve, and PF coils as square markers. 
    (b) Finite-$\beta$ QA hybrid configurations with self-consistent bootstrap current for different prescribed profile families. 
    (c) Large-$\iota$ QA and TO-like configurations used for the ideal-ballooning stability comparison.
    }
    \label{fig2}
\end{figure*}

The vacuum cross sections are shown in Fig.~\ref{fig2}(a). 
The cases VQA2-i12 and VQA2-i20 demonstrate that vacuum QA configurations with magnetic well can be obtained at different levels of external rotational transform after re-optimization of the coil currents and equilibrium parameters. 
Thus the vacuum optimization provides QA reference fields with both controllable ERT and magnetic well.

The magnetic-well objective in the vacuum cases should be interpreted as a vacuum-field shaping constraint. 
In vacuum configurations, the magnetic well is a useful indicator for avoiding unfavorable interchange instabilities. 
For finite-$\beta$ equilibria, pressure-driven stability is more appropriately assessed with the Mercier criterion and ideal-ballooning diagnostics, since the magnetic-well contribution is only one part of the full stability balance.

These vacuum results establish the baseline flexibility of the PDC framework: the optimized current distribution can generate nested QA vacuum fields with prescribed ERT and magnetic well. 
These cases serve as reference fields for the finite-$\beta$ QA hybrid configurations discussed next.

\section{Finite-$\beta$ QA hybrid configurations}
\label{sec:finite_beta}

We next construct finite-$\beta$ QA hybrid equilibria with self-consistent bootstrap current. 
The finite-$\beta$ cases in this section are based on the vacuum QA references discussed above. 
The prescribed density, temperature, and pressure profiles are treated as input profiles, while the toroidal-current-density ($dI(s)/ds$) profile is optimized to match the neoclassical bootstrap-current prediction. 
The PDC currents, PF currents, and selected equilibrium parameters are re-optimized in the free-boundary calculation.

This section considers three aspects of finite-$\beta$ operation. 
Section~\ref{sec:high_beta} compares three high-$\beta$ equilibria with different profile shapes as steady-state-relevant QA hybrid examples. 
Section~\ref{sec:beta_ramp} studies a $\beta$ ramp-up sequence using NCSX-like profiles. 
Section~\ref{sec:nfp_variation} examines field-period variation by re-optimizing the current distribution and equilibrium parameters within the same PDC framework.

\subsection{High-$\beta$ operation with different profiles}
\label{sec:high_beta}

We first compare three high-$\beta$ QA hybrid equilibria, QA2-C5, QA2-I4, and QA2-X5. 
All three cases have $n_{\rm fp}=2$ and $\langle\beta\rangle>3.5\%$, but use different prescribed profile sets. 
The labels C, I, and X denote cubic, ITER-like, and NCSX-like profiles, respectively. 
These profiles represent different possible heating and fueling conditions and therefore lead to different pressure-gradient, bootstrap-current, and rotational-transform profiles. 
For these cases, the active targets are the averaged aspect ratio, Mercier stability, bootstrap-current consistency, QA breaking, and coil-current constraints.

\begin{figure*}[htbp]
    \centering
    \includegraphics[width=1.0\linewidth]{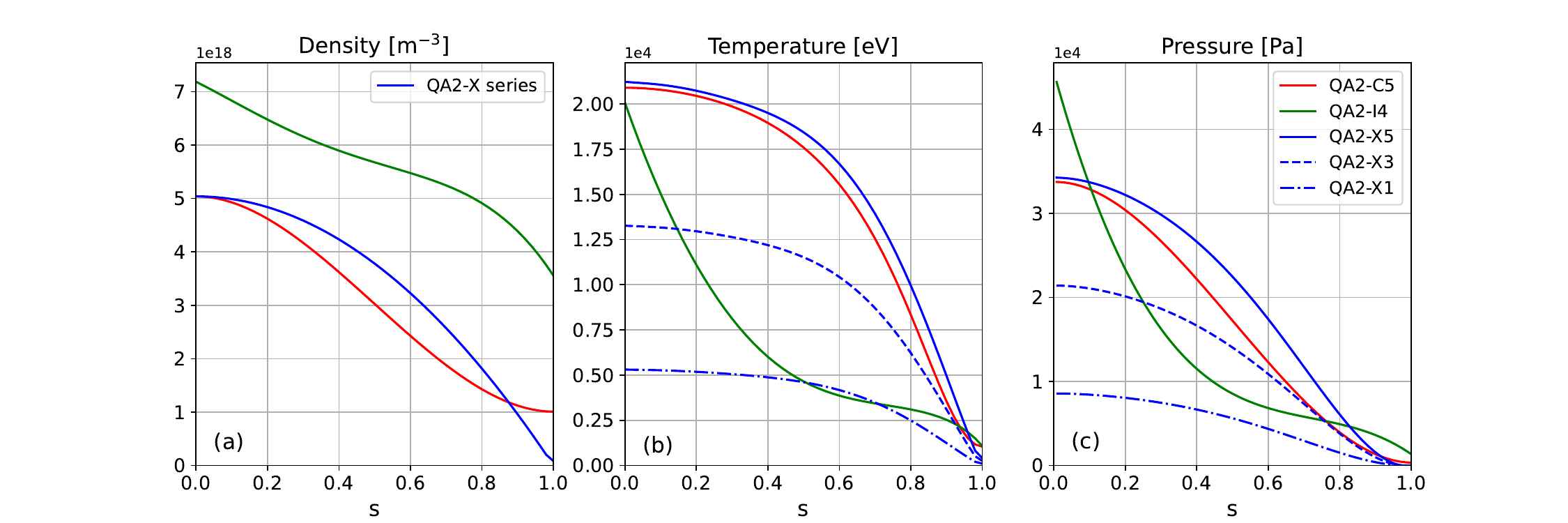}
    \caption{
    Prescribed input profiles for the finite-$\beta$ QA hybrid configurations. 
    (a) Density, (b) temperature, and (c) pressure profiles as functions of the normalized toroidal flux $s$. 
    The solid curves compare the three high-$\beta$ profile families, QA2-C5, QA2-I4, and QA2-X5. 
    The dashed and dash-dotted curves show the NCSX-like $\beta$ ramp-up sequence, QA2-X3 and QA2-X1, obtained by changing the temperature amplitude while keeping the density profile fixed.
    }
    \label{fig3}
\end{figure*}

\begin{figure*}[htbp]
    \centering
    \includegraphics[width=0.8\linewidth]{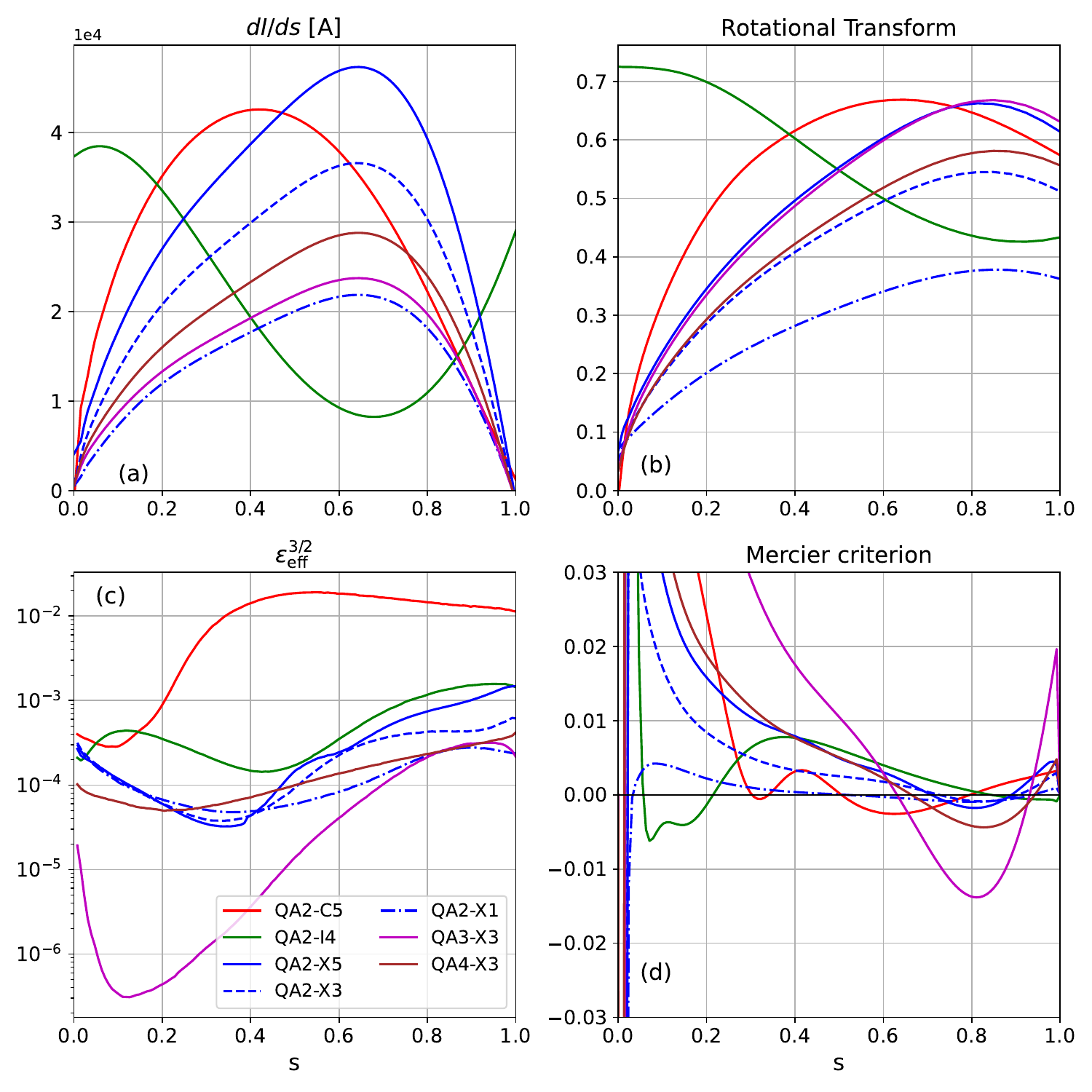}
    \caption{
    Profiles obtained from the optimized finite-$\beta$ QA hybrid equilibria. 
    (a) Toroidal-current-density profile $dI(s)/ds$, optimized to satisfy bootstrap-current consistency. 
    (b) Rotational-transform profile $\iota(s)$. 
    (c) Effective ripple metric $\epsilon_{\rm eff}^{3/2}$. 
    (d) Mercier criterion.
    The curves include the high-$\beta$ profile-family comparison, the NCSX-like $\beta$ ramp-up sequence, and the field-period-variation cases QA3-X3 and QA4-X3.
    }
    \label{fig4}
\end{figure*}

The prescribed density, temperature, and pressure profiles are shown in Fig.~\ref{fig3}(a)--(c). 
The cubic profile provides a smooth reference case. 
The ITER-like profile is more core-peaked and produces a stronger bootstrap-current drive. 
The NCSX-like profile is broader and is also used later for the $\beta$ ramp-up sequence. 
For each profile set, the current-density profile $dI(s)/ds$ is optimized through the bootstrap-current consistency objective. 
The resulting $dI(s)/ds$ profiles are shown in Fig.~\ref{fig4}(a).

The free-boundary cross sections of the three optimized equilibria are compared in Fig.~\ref{fig2}(b). 
Although the profiles are substantially different, all three cases yield QA hybrid equilibria within the same PDC framework. 
The plasma boundary and shaping respond to the pressure and current profiles through the free-boundary equilibrium, while the coil geometry itself is unchanged.

Figure~\ref{fig4}(b) shows the rotational-transform profiles of the three cases. 
The different profile sets lead to different bootstrap-current and magnetic-shear distributions, which in turn modify the rotational-transform profile. 
This dependence is important because the rotational-transform profile affects rational-surface locations, magnetic-shear structure, and pressure-driven stability properties. 
Thus, changing the profile shape is not only a change in the scalar value of $\langle\beta\rangle$, but also changes the equilibrium structure relevant to stability and transport. 

The neoclassical transport metric $\epsilon_{\rm eff}^{3/2}$ and the Mercier stability metric are shown in Fig.~\ref{fig4}(c)-(d). 
The three cases retain acceptable QA quality and Mercier stability over most of the plasma volume, although their edge behavior differs because of the coupled response of pressure, current, and boundary shape. 
These cases show that the PDC framework can support high-$\beta$ QA hybrid equilibria with self-consistent bootstrap current for multiple physically distinct profile families, providing steady-state-relevant operation modes with reduced external current-drive demand.

\subsection{$\beta$ ramp-up with NCSX-like profiles}
\label{sec:beta_ramp}

We next consider a $\beta$ ramp-up sequence based on the NCSX-like profile family. 
The three cases QA2-X1, QA2-X3, and QA2-X5 have the same field period and profile shape, with $\langle\beta\rangle\sim 1\%$, $\sim 3\%$, and $\sim 5\%$, respectively. 
In this sequence, the density profile is kept fixed, while the temperature is increased to raise the pressure and hence the volume-averaged beta. 
For each $\beta$ level, the PDC currents, PF currents, toroidal-current-density profile, and selected equilibrium parameters are re-optimized in the free-boundary calculation. 
The active targets are the averaged aspect ratio, Mercier stability, bootstrap-current consistency, QA breaking, and coil-current constraints.

This sequence is intended to mimic a gradual increase of plasma performance in an experiment rather than three unrelated equilibria. 
During such a ramp-up, it is important that the required changes in shaping-coil currents remain moderate, since large current excursions would complicate superconducting magnet operation and increase the demands on power supplies and cryogenic systems. 
In the present sequence, the optimized PDC current distributions change only modestly as $\beta$ is increased. 
This behavior is favorable for an HTS-based planar shaping-coil system, where current-ramp and transient electromagnetic effects are important engineering considerations \cite{Nash2025Canis,Abdollahi2018HTSRamp}.

The prescribed density, temperature, and pressure profiles are shown in Fig.~\ref{fig3}(a)--(c). 
Because the density profile is unchanged, the increase of pressure is mainly controlled by the temperature. 
The larger pressure gradient strengthens the bootstrap-current drive, leading to the optimized current-density profiles shown in Fig.~\ref{fig4}(a). 
The radial structure of $dI(s)/ds$ remains tied to the NCSX-like profile family, while its magnitude increases with $\beta$.

The corresponding rotational-transform profiles are shown in Fig.~\ref{fig4}(b). 
As the pressure and bootstrap current increase, the total rotational transform and magnetic shear are modified. 
This change is relevant for finite-$\beta$ operation because the rotational-transform profile influences rational-surface locations, shear structure, and pressure-driven stability. 
The ramp-up sequence demonstrates that the optimized QA hybrid can maintain favorable Mercier stability over most of the plasma volume while the pressure and bootstrap current are increased.

The neoclassical transport metric $\epsilon_{\rm eff}^{3/2}$ and Mercier criterion are also shown in Fig.~\ref{fig4}(c)-(d). 
The NCSX-like sequence maintains low $\epsilon_{\rm eff}^{3/2}$ and interchange stability over most of the plasma volume, although the edge value changes with $\beta$. 
Together with the moderate change in the required PDC currents, this indicates that PDC-supported QA hybrid configurations can be varied from low to high finite-$\beta$ without losing the basic QA character or requiring a large reconfiguration of the coil system.

\subsection{Field-period variation by re-optimization}
\label{sec:nfp_variation}

We finally consider field-period variation within the PDC optimization framework. 
Two finite-$\beta$ QA hybrid cases, QA3-X3 and QA4-X3, are constructed with $n_{\rm fp}=3$ and $n_{\rm fp}=4$, respectively. 
Both cases use the NCSX-like profile family with $\langle\beta\rangle\sim 2-3\%$. 
Compared with the baseline $n_{\rm fp}=2$ case QA2-X3, the PDC currents, PF currents, toroidal-current-density profile, and selected equilibrium parameters are re-optimized for the target field period.

This study is not intended to represent a direct switch of field period at fixed plasma shape. 
Instead, it tests whether different finite-$\beta$ QA hybrid branches can be accessed within the same planar-coil platform after re-optimization. 
The optimized aspect ratios are larger than in the baseline $n_{\rm fp}=2$ cases, with $\bar{A}\simeq 5.70$ for QA3-X3 and $\bar{A}\simeq 4.86$ for QA4-X3. 
The increased aspect ratio helps accommodate the higher field periodicity and improves the matching between the target equilibrium and the available shaping field.

\begin{figure*}[htbp]
    \centering
    \includegraphics[width=1.0\linewidth]{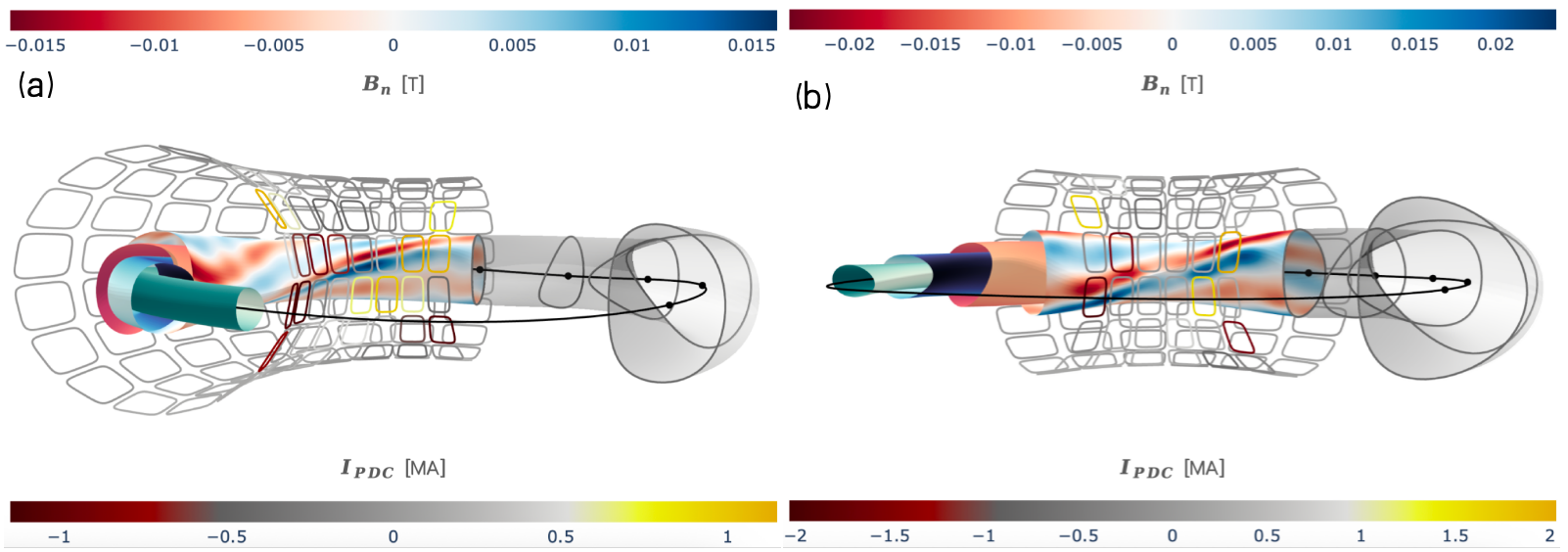}
    \caption{
    Field-period variation by re-optimization. 
    The optimized finite-$\beta$ QA hybrid configurations QA3-X3 (a) and QA4-X3 (b) are shown for $n_{\rm fp}=3$ and $n_{\rm fp}=4$, respectively. 
    Both cases use the NCSX-like profile family with $\langle\beta\rangle\sim 2-3\%$. 
    The normal magnetic field $B_{\rm n}$ on the target plasma boundary and the corresponding optimized PDC current distributions are shown. 
    The different periodic structures of $B_{\rm n}$ and the PDC currents indicate that distinct field-period branches can be generated within the same planar-coil framework after re-optimization.
    }
    \label{fig5}
\end{figure*}

Figure~\ref{fig5} shows the optimized QA3-X3 and QA4-X3 configurations. 
The normal magnetic field $B_{\rm n}$ and the PDC current distribution reorganize according to the target periodicity. 
The different periodic structures of the optimized current patterns indicate that the same planar-coil framework can generate distinct field-period branches after re-optimization.

The rotational-transform, effective-ripple and Mercier criterion profiles are included in Fig.~\ref{fig4}(b)-(d). 
The two re-optimized cases retain finite rotational transform, acceptable neoclassical confinement and Mercier stability over most of the plasma volume. 
Together with the $n_{\rm fp}=2$ sequence discussed above, these results show that the PDC framework can access multiple field-period branches of finite-$\beta$ QA hybrid equilibria, although each branch requires its own free-boundary re-optimization.

\section{TO-like large-$\iota$ configurations}
\label{sec:TO}
QA is a natural starting point for hybrid designs because it approximately preserves the favorable guiding-center properties of axisymmetric tokamaks while allowing three-dimensional shaping \cite{Zarnstorff2001,Ku2009,LandremanPaul2022}. 
Omnigenity provides a broader design space than quasisymmetry \cite{LandremanCatto2012,Dudt2024Omnigenity}, and recent work suggests that finite-$\beta$ omnigenous equilibria can be optimized for improved ballooning stability \cite{Gaur2025Omni}. 
This motivates us to explore TO-like large-$\iota$ configurations as a controlled geometry comparison. 
Two cases are considered: QA2-X3-Li and TO2-X3-Li. 
They use the same field period, comparable volume-averaged beta, and the same prescribed NCSX-like pressure and current-density profiles. 
The goal is to change the magnetic geometry from QA to TO while keeping the other equilibrium inputs as close as possible. 
For both cases, the Mercier-stability target is included so that the interchange stability is improved before applying the ideal-ballooning diagnostic. 
No ballooning-stability metric is included in the optimization objective, so the stability difference discussed below is not a direct result of ballooning optimization.

For QA2-X3-Li, the active targets are the averaged aspect ratio, Mercier stability, QA breaking, and coil-current constraints. 
For TO2-X3-Li, the QA target is replaced by the TO target, while the averaged-aspect-ratio, Mercier, and coil-current constraints are retained. 
The rotational-transform profile $\iota(s)$ is not imposed as an optimization target, but is obtained from the final free-boundary equilibria in post-processing. 
The resulting $\iota(s)$ profiles of the two cases are found to be close, so the comparison can be regarded as a matched-profile comparison between QA and TO-like magnetic geometries.

\begin{figure*}[htbp]
    \centering
    \includegraphics[width=1.0\linewidth]{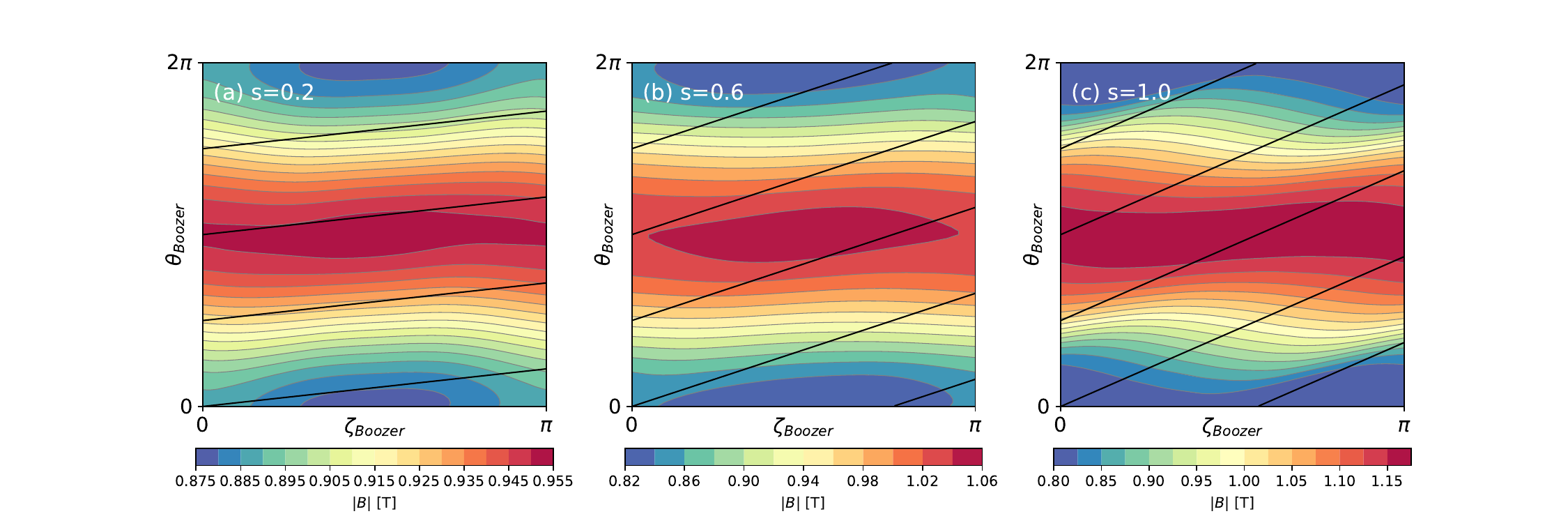}
    \caption{
    Boozer-space contours of $|B|$ for the TO2-X3-Li configuration on three representative flux surfaces: (a) $s=0.2$, (b) $s=0.6$, and (c) $s=1.0$. 
    The black solid lines denote magnetic field lines in Boozer coordinates. 
    The TO target is imposed on the outermost surface, $s=1.0$, where the TO character is most evident.
    }
    \label{fig6}
\end{figure*}

Figure~\ref{fig2}(c) compares the plasma cross sections of the two large-$\iota$ cases. 
Figure~\ref{fig6} shows the Boozer-space structure of $|B|$ for the TO2-X3-Li case on three different flux surfaces. 
Since the TO condition is defined on each flux surface, the present optimization imposes the TO target only on the LCFS. 
The inner flux surfaces are obtained self-consistently from the free-boundary equilibrium rather than optimized independently for TO. 
The Boozer contours show that the TO character is most evident near the optimized outermost surface.

\begin{figure*}[htbp]
    \centering
    \includegraphics[width=1.0\linewidth]{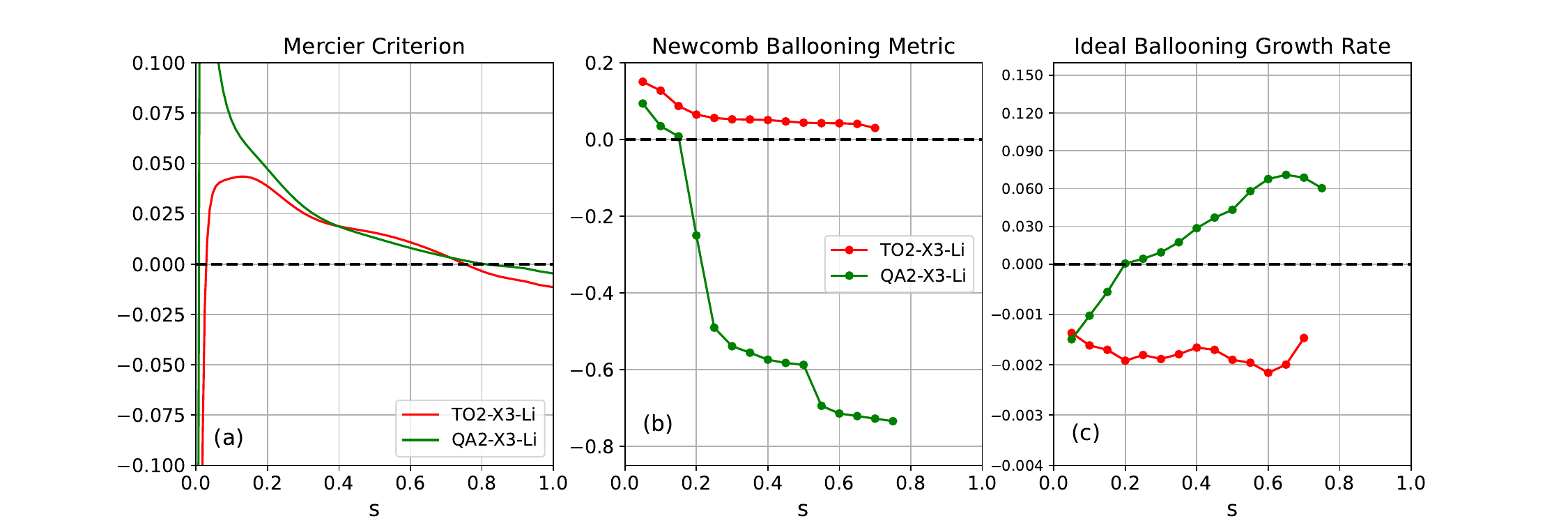}
    \caption{
    Local pressure-driven stability comparison for the large-$\iota$ QA and TO-like configurations. 
    (a) Mercier criterion, (b) Newcomb ballooning metric, and (c) infinite-$n$ ideal-ballooning growth rate. 
    The dashed horizontal line marks the marginal-stability boundary. 
    The TO2-X3-Li case remains locally stable to the evaluated ideal-ballooning modes, while the QA2-X3-Li reference becomes unstable in the corresponding radial range.
    }
    \label{fig7}
\end{figure*}

After optimization, the pressure-driven stability is evaluated only in the Mercier-stable regions. 
We use two ballooning-related diagnostics: the Newcomb ballooning metric and the standard infinite-$n$ ideal-ballooning eigenvalue calculation \cite{Mercier1964,Newcomb1960,Dudt2020DESC,Sanchez2000BOOZ,Gaur2025Omni}. 
The Newcomb metric gives a necessary condition for ideal-ballooning stability along magnetic field lines, whereas the eigenvalue calculation provides the final ballooning stability diagnostic. 
The two diagnostics use opposite sign conventions. 
For the Newcomb metric shown in Fig.~\ref{fig7}(b), positive values indicate that the Newcomb necessary condition is satisfied, while negative values indicate violation of this condition. 
For the ideal-ballooning eigenvalue shown in Fig.~\ref{fig7}(c), positive values correspond to unstable ideal-ballooning modes, while negative values indicate stability. 

The stability results are shown in Fig.~\ref{fig7}. 
Figure~\ref{fig7}(a) shows that both cases are Mercier stable over most of the plasma volume. 
However, the two cases behave differently in the subsequent ballooning diagnostics. 
As shown in Fig.~\ref{fig7}(b), the Newcomb metric becomes negative for QA2-X3-Li over part of the evaluated region, indicating violation of the Newcomb necessary condition, whereas it remains positive for TO2-X3-Li. 
Consistently, Fig.~\ref{fig7}(c) shows positive ideal-ballooning eigenvalues for QA2-X3-Li and negative eigenvalues for TO2-X3-Li. 
Thus, after local interchange stability has been addressed through the Mercier-stability optimization, the TO-like configuration remains locally stable to the evaluated ideal-ballooning modes, while the QA reference becomes unstable.

The other finite-$\beta$ QA hybrid configurations discussed in Sec.~\ref{sec:finite_beta} are also found to be locally unstable to the same ideal-ballooning diagnostic. 
That is, among the finite-$\beta$ configurations examined in this work, only the TO-like case remains stable in the local infinite-n ideal-ballooning analysis. 
This comparison suggests that moving from QA toward TO can provide additional geometric freedom for improving pressure-driven stability at large rotational transform.

This result should be interpreted as a stability-oriented geometry exploration, not as a complete steady-state TO hybrid demonstration. 
The finite-$\beta$ QA hybrids in Sec.~\ref{sec:finite_beta} establish bootstrap-current-consistent operation in the present framework. 
The TO comparison shows a complementary point: because TO is less restrictive than QA and allows a broader class of three-dimensional $|B|$ structures, it may provide additional freedom for improving ideal-ballooning stability after interchange stability has been addressed. 
Developing a fully self-consistent TO hybrid configuration that simultaneously satisfies bootstrap-current consistency, Mercier stability, ballooning stability, and coil-current limits remains a natural direction for future work.

\section{Summary and discussion}
\label{sec:summary}

In this work, we use a single-stage free-boundary optimization framework to obtain finite-$\beta$ tokamak--stellarator hybrid configurations supported by PDCs. 
The PDCs are arranged on an axisymmetric winding surface with fixed positions, while the PDC currents, PF currents, and selected plasma-equilibrium parameters are optimized simultaneously. 
This framework provides a flexible way to generate different three-dimensional shaping fields without redesigning the coil geometry.
We first demonstrate vacuum QA configurations with controllable external rotational transform and magnetic well. 
The programmed PDC current distribution produces nested QA vacuum fields and provides reference configurations for finite-$\beta$ optimization. 

The main finite-$\beta$ result is the construction of QA hybrid equilibria with self-consistent bootstrap current and favorable Mercier stability. 
For cubic, ITER-like, and NCSX-like profile families, the optimized equilibria maintain QA quality while matching the neoclassical bootstrap-current prediction. 
These results show that the PDC framework can support steady-state-relevant QA hybrid equilibria with substantial bootstrap-current contribution and reduced external current-drive demand.
The NCSX-like $\beta$ ramp-up sequence shows that low-, intermediate-, and high-$\beta$ equilibria can be obtained within the same profile family by increasing the temperature amplitude at fixed density and re-optimizing the equilibrium. 
The required changes in the optimized PDC currents remain moderate during this process, which is favorable for superconducting shaping-coil operation. 
Re-optimized finite-$\beta$ QA hybrid configurations with $n_{\rm fp}=3$ and $n_{\rm fp}=4$ further show that different field-period branches can be accessed within the same PDC framework.

At large rotational transform, both the QA reference and TO-like configurations are further optimized with a Mercier-stability target. 
The TO-like configuration shows infinite-$n$ ideal-ballooning stability without directly optimizing a ballooning metric, whereas the QA reference and the finite-$\beta$ QA hybrids considered in Sec.~\ref{sec:finite_beta} are locally unstable to the same diagnostic. 
This result suggests that TO-like shaping can provide additional freedom for pressure-driven stability, although the present TO case should be viewed as a stability-oriented extension rather than a fully bootstrap-current-consistent operating point.

Beyond the equilibrium optimization demonstrated here, the current-programmable PDC system may also be useful for edge and control applications. 
Programmable three-dimensional fields could be used to explore island-divertor or edge-topology control, resonant magnetic perturbation spectra for ELM control, and feedback control of magnetic shape or field errors \cite{Feng2006IslandDivertor,Evans2006RMP,Degrave2022MagneticControl,Nash2025Canis}. 
These applications are not pursued in the present work and would require dedicated plasma-response, scrape-off-layer, and control-system modeling, but they provide additional motivation for a configurable PDC platform.

The optimized PDC currents also indicate important engineering issues. 
The averaged PDC current is relatively modest, but the peak current can be large and is mainly concentrated on a subset of high-field-side coils. 
A complete engineering assessment will require further optimization of current density, ampere-turns, electromagnetic forces, structural support, shielding, cooling, and integration with surrounding tokamak components.

Overall, these results show that PDC-based hybrids provide a configurable free-boundary platform for finite-$\beta$ QA equilibria and TO-oriented stability exploration. 
Future work should combine bootstrap-current consistency, TO quality, Mercier and ballooning stability, and more realistic coil-engineering constraints in a single optimization loop. 
Further studies of global MHD stability, resistive modes, energetic-particle confinement, edge-field control, and active feedback will also be needed before reactor-level applicability can be assessed.

\ack
We thank Y.B. Li for providing relevant data of J-TEXT. YL, ZZ, and YZ are supported by the National MCF R\&D Program under Grant Number 2024YFE03230400, the National Natural Science Foundation of China under Grant Number 12305246, and the Fundamental Research Funds for the Central
Universities. 
HL, GY, and CZ are supported by the National Natural Science Foundation of China under Grant No. 12405267, the Strategic Priority Research Program of the Chinese Academy of Sciences under Grant No. XDB0790302.
This work is also supported by the
National Key Laboratory of Frontier Physics in Controlled Fusion.

\section*{References}
\bibliographystyle{unsrt}
\bibliography{ref}

\end{document}